# LLMs can generate robotic scripts from goal-oriented instructions in biological laboratory automation

Takashi Inagaki[1,2,*], Akari Kato[1,3], Koichi Takahashi[1,3,4,5], Haruka Ozaki[1,6,7,*], Genki N. Kanda[1,3,*]

[1] Laboratory Automation Supplier's Association, Japan
[2] Department of Mechanical Systems Engineering, Graduate School of Engineering, Nagoya University, Nagoya, Aichi, Japan
[3] Laboratory for Biologically Inspired Computing, RIKEN Center for Biosystems Dynamics Research, Kobe, Hyogo, Japan
[4] Graduate School of Media and Governance, Keio University, Fujisawa, Japan
[5] Graduate School of Frontier Biosciences, Osaka University, Suita, Japan
[6] Bioinformatics Laboratory, Institute of Medicine, University of Tsukuba, Tsukuba, Ibaraki, Japan
[7] Center for Artificial Intelligence Research, University of Tsukuba, Tsukuba, Ibaraki, Japan

**\* Contact authors:**
Takashi Inagaki            inagaki.takashi.n7@s.mail.nagoya-u.ac.jp
Haruka Ozaki, Ph.D.        haruka.ozaki@md.tsukuba.ac.jp
Genki N. Kanda, Ph.D., PMP genki.kanda@riken.jp

## Abstract

The use of laboratory automation by all researchers may substantially accelerate scientific activities by humans, including those in the life sciences. However, computer programs to operate robots should be written to implement laboratory automation, which requires technical knowledge and skills that may not be part of a researcher's training or expertise. In the last few years, there has been remarkable development in large language models (LLMs) such as GPT-4, which can generate computer codes based on natural language instructions. In this study, we used LLMs, including GPT-4, to generate scripts for robot operations in biological experiments based on ambiguous instructions. GPT-4 successfully generates scripts for OT-2, an automated liquid-handling robot, from simple instructions in natural language without specifying the robotic actions. Conventionally, translating the nuances of biological experiments into low-level robot actions requires researchers to understand both biology and robotics, imagine robot actions, and write robotic scripts. Our results showed that GPT-4 can connect the context of biological experiments with robot operation through simple prompts with expert-level contextual understanding and inherent knowledge. Replacing robot script programming, which is a tedious task for biological researchers, with natural-language LLM instructions that do not consider robot behavior significantly increases the number of researchers who can benefit from automating biological experiments.

## Keywords

## Introduction

Laboratory automation is a powerful tool to enhance the efficiency and reproducibility of biological experiments. Many experimental automation systems for life sciences, including molecular biology, cell biology, and systems biology, have been developed and used in research settings till date[1–5]. However, such systems are generally not operable in a natural language, and mastering their operating language, often in a domain-specific programming language, is an obstacle to widespread laboratory automation[6].

In the past few years, large language models (LLMs), such as GPT-4, which can generate computer codes from natural language instructions, have made remarkable progress. Attempts have been made to use LLMs for directly generating program codes to control robots using natural language prompts. In organic chemistry, an AI/robotic system containing LLM has successfully and autonomously performed catalytic cross-coupling reactions using natural language inputs[7]. Similarly, in the life sciences, if LLMs can simplify the translation of experimental processes and instructions into precise robotic movements, then the cost of educating life science researchers to use robots can be substantially reduced.

In this study, we constructed a process for converting human-written natural language descriptions into robot executable scripts for an open-source liquid-handling machine (OT-2, Opentrons Labworks Inc.)[8–10] with a well-documented Python API, and evaluated how well LLMs, including GPT-4, can perform such tasks. Our results showed that GPT-4 could successfully generate robotic scripts, even from ambiguous instructions, similar to a professor's guidance for graduate students in biology laboratories.

## Methods

We developed a pipeline using GPT-4, which generates a Python script for OT-2 from the natural language description of a sequence of experimental procedures. Specifically, we implemented an architecture (**Figure 1**) that leverages the OpenAI API to generate OT-2-compatible Python scripts using LLMs based on the provided prompts. The architecture consists of (0) querying a prompt to an LLM via the OpenAI API; (1) extracting a Python script from the outputs of the LLM; (2) validating the scripts using opentrons_simulate, a simulator for OT-2 Python scripts, and storing the output results; and (3) extracting and querying the error message to the language model for correction; otherwise deeming the validation result as a success. This loop of (1) to (3) was repeated until either no error was raised or a pre-determined number of iterations (five in our case) was reached. Python version 3.8.12 was used for the implementation, and the OpenAI API was used from 2023-03-20 to 2023-04-03 (see **Appendix 1** for details).

For validating the generated Python scripts, we defined the script as a "Success" if it completes successfully and an "Error" when executed with opentrons_simulate (version 6.2.1). The errors were categorized into Opentrons error (including ApiDeprecationError, ExceptionInProtocolError, and MalformedProtocolError) and Python error (FileNotFoundError, IndentationError, NameError, RuntimeError, and SyntaxError) (see **Appendix 2** for details). If the LLM API was in a busy state and there was no response, the session was terminated and labeled as "LLM API busy".

We tested the performance of the proposed architecture in generating OT-2 Python scripts using a cell culture medium exchange protocol (such as human mesenchymal stem cells, MSCs) under various conditions, such as different language model settings (gpt-3-turbo and gpt-4 in this study) and varying numbers of iterations (loops). By examining the success rate and error types under these diverse circumstances, we aimed to evaluate the robustness and adaptability of our GPT-4-based architecture for generating Python scripts for biological laboratory automation.

## Results

We conducted validation tests using two different methods of providing instructions with different levels of detail and ambiguity. First way was 'step-by-step instruction' where the user explicitly describes procedures to execute (such as pipette xx mL reagent A to a culture dish, and then wait for yy minutes) and specifies the details of available labware (such as corning_6_wellplate_16.8ml_flat). The second way was 'goal-oriented instruction' where the user describes the goal of the experimental operation (such as exchange cell culture medium) and conveys more general information on available labware (such as just '6 well plate').

While providing the step-by-step instructions for cell culture medium exchange (**Figure 2A**), gpt-3-turbo failed in all 10 independent sessions in the first iteration, but a cumulative total of five sessions were successful, with 50% success rate after five iterations (**Figure 2B**, **C**, **Appendix 3**). However, out of the 22 independent sessions with gpt-4, 13 were successful in the first iteration (success rate 59%). After five iterations, a cumulative total of 21 sessions were successful (success rate 95%; excluding aborted sessions due to LLM API busy from the total, 100%).

Next, we examined the performance of the proposed system with goal-oriented and more ambiguous instructions. **Figure 2D** shows the prompt used, which is less specific than the step-by-step instructions. This may convey information similar to that communicated when a senior scientist asks a graduate student to perform a medium exchange for cell culture in a biology laboratory. None of 12 independent sessions using gpt-3-turbo were successful in the first iteration, and a cumulative total of two sessions were successful (success rate 17% up to five iterations; **Figure 2E**, **Appendix 3**). In contrast, out of the 17 independent sessions with gpt-4, 3 were successful in the first iteration, and a cumulative total of 12 successes were achieved, with a success rate of 71% (excluding aborted sessions due to LLM API busy from the total, 92%) when up to 5 iterations were performed when up to five iterations were performed (**Figure 2E**). These results suggest that the proposed architecture can achieve a high success rate, even when given less explicit prompts. Interestingly, the script generated by gpt-4 utilized PBS as the solution to wash the cells and used D-MEM as a fresh medium, despite the lack of mention of the role of PBS and D-MEM in this prompt (**Figure 2F**). In addition, the scripts provided biologically acceptable estimates of specific liquid volumes even when not explicitly stated in the prompt.

## Discussion

In this study, we demonstrated that GPT-4 generates functional scripts when the user provides detailed step-by-step instructions for the experimental procedures or goal-oriented instructions that tend to be substantially less explicit and ambiguous. The results suggest that the advanced language model can effectively interpret and execute tasks similar to graduate students working under the guidance of a professor. The successful generation of Python scripts for the OT-2 liquid-handling robot highlighted the potential of GPT-4 for streamlining complex processes in biological research. Furthermore, the ability to handle indirect instructions given in natural language may open new avenues for automating experimental procedures, enabling researchers to focus on result interpretation and analysis, thereby accelerating scientific discovery.

Despite the reported successful attempts, generating Python scripts based on longer instructions remains challenging. We created Python scripts using longer natural language descriptions as prompts. Specifically, we used the natural language description found in the "Protocol Steps" section of the "Cell Viability and Cytotoxicity Assay" listed on the Opentrons Protocol Library (https://protocols.opentrons.com/). Although the resulting script appeared to run without errors and was deemed successful, a close examination revealed that the generated script merely circumvented errors without executing the experimental operations as instructed. This suggests that further refinements, such as modularization using functions, may be necessary to generate scripts for complex experimental protocols involving numerous steps. Likewise, in this paper, we only showed the results of a single experimental protocol to demonstrate the potential use of LLM for biological laboratory automation, and it remains to be investigated whether the findings can be generalized to other protocols and experimental parameters in future research.

In this study, we generated Python scripts for Opentrons OT-2. Extrapolating the proposed architecture to other laboratory automation robots may present certain challenges. Python is the second most-used programming language worldwide (https://octoverse.github.com/2022/top-programming-

languages), and documentation and example scripts for the Opentrons Python API are publicly available on the vendor's website. However, other laboratory automation robots may employ different programming languages such as C# or may not have publicly available documentation. Therefore, to achieve a success similar to that of other automated laboratory robots, further validation and refinement may be necessary.

Replacing the generation of robot scripts with programming languages, which is a tedious task for biological researchers, with natural-language LLM instructions that do not consider robot behavior may significantly increase the number of researchers who can benefit from automating biological experiments.

## Funding
This study was supported by the JST-Mirai Program (JPMJMI20G7, JST, to **Takahashi K** and **Ozaki H**).

## Author Contributions
**Inagaki T**, **Kato A**, **Ozaki H**, and **Kanda GN** designed the study. **Inagaki T** developed the software. **Kanda GN** managed the project. **Takahashi K** supervised the study. All the authors discussed the results and wrote the manuscript.

## Declaration of Interests
All authors declare no competing interests.

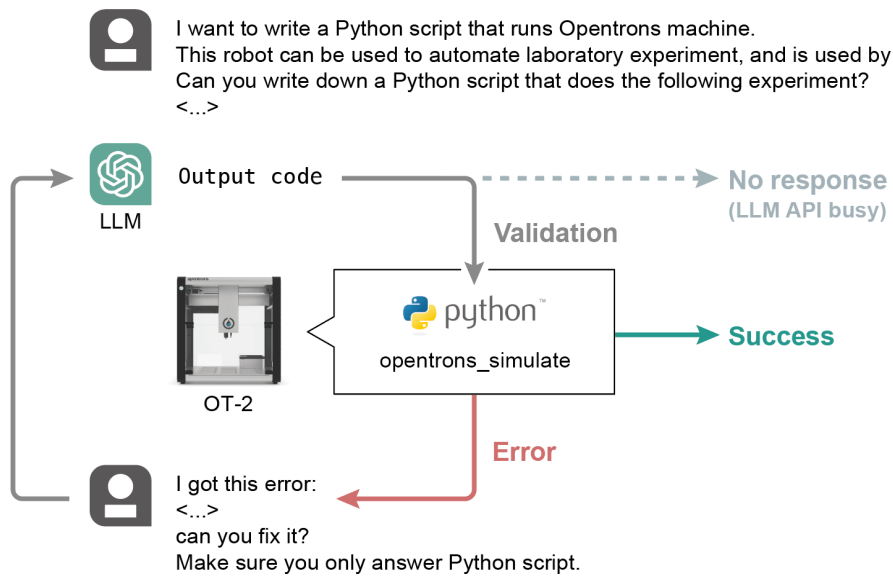

**Figure 1: Architecture overview**
Prompts were provided to the large language model (LLM) through the OpenAI API to generate Python scripts for OT-2. The scripts were extracted from the LLM-generated responses and validated using opentrons_simulate. If script validation fails, the obtained error message is used as a new prompt for the LLM, requesting a corrected script. This process was repeated a maximum of five times. The procedure is terminated after a successful attempt or once the number of iterations exceeds five. Credits: OT-2 image, https://opentrons.com/products/robots/ot-2/; ChatGPT icon image, https://openai.com/.

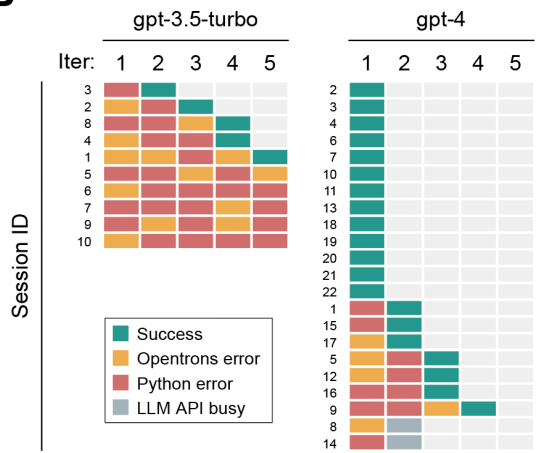
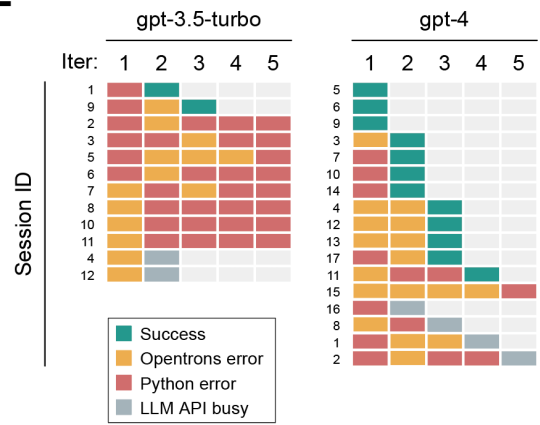

**Figure 2: Evaluation of the proposed architecture for generating Python scripts for operating OT-2**
The prompts describing instructions for cell medium exchange, examples of Python scripts obtained from prepared prompts, and the results of validating these scripts using opentrons_simulate for 'step-by-step instruction' (**A–C**) and 'goal-oriented instruction' (**D–F**).
(**A** and **D**) Prompt instructions for the cell medium exchange.
(**B** and **E**) Counts of success and error outcomes resulting from providing this prompt to LLMs in each iteration and session. Success (green), Opentrons error (yellow), Python error (red), LLM API busy (gray).
(**C** and **F**) Representative successful Python scripts generated by GPT-4 in response to prompts.

# Appendix

**Appendix 1: Messages parameter in OpenAI API calls**

The messages parameters in OpenAI API calls are as follows:

```
messages=[
    {"role": "system", "content": "You are a helpful assistant who is
    an expert in biology, computer science, and engineering."},
    {"role": "user", "content": {{prompt}}},
]
```

The source codes and result texts are available at https://github.com/labauto/Inagaki_2023_GPT4OT2.

**Appendix 2: Opentrons error and Python error**

**Opentrons error**
- ApiDeprecationError: This error occurs when the Python API version and the robot server version specified in the script do not match. Specific to the Opentrons Python API.
- ExceptionInProtocolError: This error occurs when the program itself is correct, but the experimentation protocol is infeasible. Specific to the Opentrons Python API.
- MalformedProtocolError: This error occurs when a program does not follow the Opentrons script. Specific to the Opentrons Python API.

**Python error**
- FileNotFoundError: This error occurs when a file or directory is requested but does not exist.
- IndentationError: This error occurs when an incorrect indentation exists in the source code.
- NameError: This error occurs when a variable or function is referenced before it is defined or initialized.
- RuntimeError: This error occurs when an error that does not fall into any of the other categories is detected.
- SyntaxError: This error occurs when the parser encounters syntax in source code.

**Appendix 3: Validation result of Python scripts generated by LLMs for each session and iteration**

The success and error counts of the opentrons_simulate validation for Python scripts generated by LLMs (gpt-3.5-turbo and gpt-4) were shown separately for two types of prompts for each session and iterations. The errors were divided into two categories: Python error (red), which occurs when the Python execution fails; Opentrons error (yellow), which occurs during validation by opentrons_simulate despite successful Python execution. LLM API busy; session terminated because the LLM API is in a busy state and there is no response.

| Prompt | Figure | Model | Session ID | Iteration 1 | Iteration 2 | Iteration 3 | Iteration 4 | Iteration 5 |
|---|---|---|---|---|---|---|---|---|
| step-by-step | 2A–C | gpt-3.5-turbo | 1 | ApiDeprecationError | ExceptionInProtocolError | RuntimeError | ExceptionInProtocolError | Success |
| step-by-step | 2A–C | gpt-3.5-turbo | 2 | ApiDeprecationError | SyntaxError | Success | | |
| step-by-step | 2A–C | gpt-3.5-turbo | 3 | SyntaxError | Success | | | |
| step-by-step | 2A–C | gpt-3.5-turbo | 4 | ApiDeprecationError | SyntaxError | RuntimeError | Success | |
| step-by-step | 2A–C | gpt-3.5-turbo | 5 | NameError | SyntaxError | ExceptionInProtocolError | RuntimeError | MalformedProtocolError |
| step-by-step | 2A–C | gpt-3.5-turbo | 6 | ExceptionInProtocolError | SyntaxError | IndentationError | SyntaxError | RuntimeError |
| step-by-step | 2A–C | gpt-3.5-turbo | 7 | SyntaxError | SyntaxError | SyntaxError | MalformedProtocolError | SyntaxError |
| step-by-step | 2A–C | gpt-3.5-turbo | 8 | SyntaxError | SyntaxError | ApiDeprecationError | Success | |
| step-by-step | 2A–C | gpt-3.5-turbo | 9 | SyntaxError | ApiDeprecationError | SyntaxError | ExceptionInProtocolError | SyntaxError |
| step-by-step | 2A–C | gpt-3.5-turbo | 10 | ApiDeprecationError | SyntaxError | SyntaxError | SyntaxError | SyntaxError |
| step-by-step | 2A–C | gpt-4 | 1 | FileNotFoundError | Success | | | |
| step-by-step | 2A–C | gpt-4 | 2 | Success | | | | |
| step-by-step | 2A–C | gpt-4 | 3 | Success | | | | |
| step-by-step | 2A–C | gpt-4 | 4 | Success | | | | |
| step-by-step | 2A–C | gpt-4 | 5 | ExceptionInProtocolError | FileNotFoundError | Success | | |
| step-by-step | 2A–C | gpt-4 | 6 | Success | | | | |
| step-by-step | 2A–C | gpt-4 | 7 | Success | | | | |
| step-by-step | 2A–C | gpt-4 | 8 | ExceptionInProtocolError | LLM API busy | | | |
| step-by-step | 2A–C | gpt-4 | 9 | FileNotFoundError | FileNotFoundError | ExceptionInProtocolError | | |
| step-by-step | 2A–C | gpt-4 | 10 | Success | | | | |
| step-by-step | 2A–C | gpt-4 | 11 | Success | | | | |
| step-by-step | 2A–C | gpt-4 | 12 | ExceptionInProtocolError | FileNotFoundError | Success | | |
| step-by-step | 2A–C | gpt-4 | 13 | Success | | | | |
| step-by-step | 2A–C | gpt-4 | 14 | FileNotFoundError | LLM API busy | | | |
| step-by-step | 2A–C | gpt-4 | 15 | FileNotFoundError | Success | | | |
| step-by-step | 2A–C | gpt-4 | 16 | SyntaxError | FileNotFoundError | Success | | |
| step-by-step | 2A–C | gpt-4 | 17 | ExceptionInProtocolError | Success | | | |
| step-by-step | 2A–C | gpt-4 | 18 | Success | | | | |
| step-by-step | 2A–C | gpt-4 | 19 | Success | | | | |
| step-by-step | 2A–C | gpt-4 | 20 | Success | | | | |
| step-by-step | 2A–C | gpt-4 | 21 | Success | | | | |
| step-by-step | 2A–C | gpt-4 | 22 | Success | | | | |
| goal-oriented | 2D–F | gpt-3.5-turbo | 1 | SyntaxError | Success | | | |
| goal-oriented | 2D–F | gpt-3.5-turbo | 2 | SyntaxError | ExceptionInProtocolError | SyntaxError | SyntaxError | SyntaxError |
| goal-oriented | 2D–F | gpt-3.5-turbo | 3 | SyntaxError | RuntimeError | MalformedProtocolError | FileNotFoundError | RuntimeError |
| goal-oriented | 2D–F | gpt-3.5-turbo | 4 | ApiDeprecationError | LLM API busy | | | |
| goal-oriented | 2D–F | gpt-3.5-turbo | 5 | RuntimeError | MalformedProtocolError | ExceptionInProtocolError | MalformedProtocolError | FileNotFoundError |
| goal-oriented | 2D–F | gpt-3.5-turbo | 6 | SyntaxError | ApiDeprecationError | SyntaxError | SyntaxError | SyntaxError |
| goal-oriented | 2D–F | gpt-3.5-turbo | 7 | ExceptionInProtocolError | SyntaxError | ExceptionInProtocolError | RuntimeError | SyntaxError |
| goal-oriented | 2D–F | gpt-3.5-turbo | 8 | ApiDeprecationError | SyntaxError | SyntaxError | SyntaxError | SyntaxError |
| goal-oriented | 2D–F | gpt-3.5-turbo | 9 | RuntimeError | MalformedProtocolError | Success | | |
| goal-oriented | 2D–F | gpt-3.5-turbo | 10 | ApiDeprecationError | SyntaxError | SyntaxError | SyntaxError | SyntaxError |
| goal-oriented | 2D–F | gpt-3.5-turbo | 11 | ApiDeprecationError | SyntaxError | SyntaxError | SyntaxError | SyntaxError |
| goal-oriented | 2D–F | gpt-3.5-turbo | 12 | ApiDeprecationError | LLM API busy | | | |
| goal-oriented | 2D–F | gpt-4 | 1 | RuntimeError | ExceptionInProtocolError | ExceptionInProtocolError | LLM API busy | |
| goal-oriented | 2D–F | gpt-4 | 2 | RuntimeError | ExceptionInProtocolError | FileNotFoundError | FileNotFoundError | LLM API busy |
| goal-oriented | 2D–F | gpt-4 | 3 | ExceptionInProtocolError | Success | | | |
| goal-oriented | 2D–F | gpt-4 | 4 | MalformedProtocolError | ExceptionInProtocolError | Success | | |
| goal-oriented | 2D–F | gpt-4 | 5 | Success | | | | |
| goal-oriented | 2D–F | gpt-4 | 6 | Success | | | | |
| goal-oriented | 2D–F | gpt-4 | 7 | FileNotFoundError | Success | | | |
| goal-oriented | 2D–F | gpt-4 | 8 | ExceptionInProtocolError | FileNotFoundError | LLM API busy | | |
| goal-oriented | 2D–F | gpt-4 | 9 | Success | | | | |
| goal-oriented | 2D–F | gpt-4 | 10 | RuntimeError | Success | | | |
| goal-oriented | 2D–F | gpt-4 | 11 | ExceptionInProtocolError | SyntaxError | SyntaxError | Success | |
| goal-oriented | 2D–F | gpt-4 | 12 | ApiDeprecationError | ExceptionInProtocolError | Success | | |
| goal-oriented | 2D–F | gpt-4 | 13 | ExceptionInProtocolError | ExceptionInProtocolError | Success | | |
| goal-oriented | 2D–F | gpt-4 | 14 | SyntaxError | Success | | | |
| goal-oriented | 2D–F | gpt-4 | 15 | ExceptionInProtocolError | ExceptionInProtocolError | ExceptionInProtocolError | ExceptionInProtocolError | FileNotFoundError |
| goal-oriented | 2D–F | gpt-4 | 16 | SyntaxError | LLM API busy | | | |
| goal-oriented | 2D–F | gpt-4 | 17 | FileNotFoundError | ExceptionInProtocolError | Success | | |